%% file: spec_hsaio.tex
\documentclass[twoside]{sfm}

\input{sf.def}

\begin{document}

\def\llm{{\sc LLmodels}}
\def\atl{{\sc ATLAS9}}
\def\aatl{{\sc ATLAS12}}
\def\starsp{{\sc STARSP}}
\def\aur{$\Theta$~Aur}
\def\logg{\log g}
\def\tauros{\tau_{\rm Ross}}
\def\kms{km\,s$^{-1}$}
\def\bz{$\langle B_{\rm z} \rangle$}
\def\degr{^\circ}
% journals
\def\aaps{A\&AS}
\def\aap{A\&A}
\def\apjs{ApJS}
\def\apj{ApJ}
\def\rmxaa{Rev. Mexicana Astron. Astrofis.}
\def\mnras{MNRAS}
\def\actaa{Acta Astron.}
\newcommand{\Tef}{T$_{\rm eff}$~}
\newcommand{\Vt}{$V_t$}
\newcommand{\CC}{$^{12}$C/$^{13}$C~}
\newcommand{\CDC}{$^{12}$C/$^{13}$C~}

\input{hsaio/ABvar.tex}

\end{document}

%% file: hsaio/ABvar.tex
\pagebreak

\thispagestyle{titlehead}

\setcounter{section}{0}
\setcounter{figure}{0}
\setcounter{table}{0}

%%%%%%%%%%%%%%%%%%%%%%%%%%%
% !!!
\markboth{Saio}{A and B type variables}

\titl{A- and B-type star pulsations in the Kepler and CoRoT era: theoretical considerations}{Saio H.}
{Institute of Astronomy, Graduate School of Science, Tohoku University, email: {\tt saio@astr.tohoku.ac.jp} }

\abstre{
Among A-type main-sequence variables, pulsations of $\delta$ Sct and $\gamma$ Dor variables are driven in the He II ionization zone, while H-ionization zone and strong magnetic fields seem to play roles in the excitation of high-order p-modes in rapidly oscillating Ap (roAp) stars. Pulsations in B-type variables, $\beta$ Cephei and slowly pulsating B (SPB) stars are excited by the $\kappa$-mechanism at the Fe-opacity bump at $T\approx 2\times10^5$K. In addition, the strange-mode instability seems responsible for the excitation of pulsations in luminous AB-supergiants ($\alpha$ Cygni variables). We discuss excitation mechanisms for pulsations in A- and B-type variables stars. 
}

\baselineskip 12pt

\section{Introduction}\label{sec:intro}

\begin{figure}[!t]
\begin{center}
\includegraphics[width=0.6\textwidth]{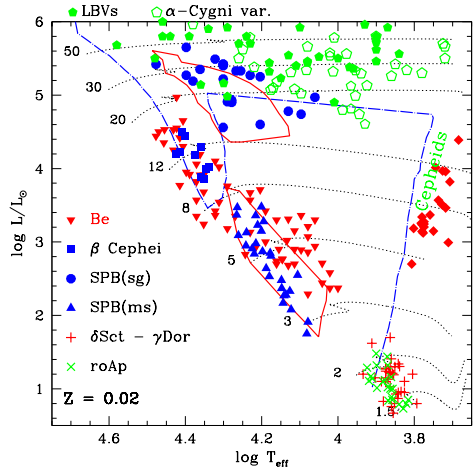}
\vspace{-5mm}
\caption[]{Some of the variable A- and B-type stars
are plotted (some classical Cepehids are also shown for comparison) in the HR diagram. The boundaries of  pulsational stability obtained for models with a standard heavy-element abundance of $Z=0.02$ are shown (based on Saio\cite{sai11}) by solid lines for g-modes, and  by a dash-dotted line for radial (p-) modes. Nearly straight part extending from $(\log T_{\rm eff},\log L/L_\odot) \approx (3.9,1.0)$ to $\approx(3.75,4.7)$ corresponds to the blue-edge of the Cepheid instability strip (red-edge is not shown). Also plotted are evolutionary tracks without core-overshooting (dotted lines) from ZAMS, along which the initial masses in solar-mass units are indicated.}
\label{fig:allvar}
\end{center}
\end{figure}

Figure\,\ref{fig:allvar} is a HR diagram showing positions of  variable A- and B-type stars, and theoretical excitation boundaries for radial (p-) modes and g-modes.  Pulsations of these stars are believed to be self excited by thermal processes, in which radiative flux plays a dominant role. 
(We do not discuss stochastic excitation by turbulent convection in this paper.) 

If the energy flux from stellar interior is blocked (released) and matter gains (loses) energy in the compressed (expanded) phase, the next expansion (contraction) will be slightly stronger than the previous one; i.e., amplitude will grow slightly; this is a simplified interpretation of the $\kappa$ (or opacity)-mechanism (e.g., Cox\cite{cox74}, Aerts et al.\cite{aer10} for detail). 
The $\kappa$-mechanism driving (in weakly nonadiabatic conditions) occurs if 
\begin{equation}
{d\over dr}\left(\kappa_T + {\kappa_\rho\over \Gamma_3 - 1}\right) > 0
\end{equation} 
(Unno et al.\cite{unn89}, ch. 5), where $\kappa_T=(\partial\ln\kappa/\partial\ln T)_\rho$,
$\kappa_\rho=(\partial\ln\kappa/\partial\ln\rho)_T$, and $\Gamma_3-1=(\partial\ln T/\partial\ln\rho)_s$ with $T$, $\rho$ and $s$ being temperature, density, and specific entropy, respectively.
The condition is satisfied only in a narrow zone of the hotter side of an opacity peak, and surrounding layers work to damp pulsations. If the driving effect exceeds the damping effect, the pulsation will be excited.
The thermal coupling in a layer is most effective if the local thermal time is comparable to the pulsation period. 
Since the thermal time is shorter if the layer is closer to the surface,  the $\kappa$-mechanism is strongest in models having $T_{\rm eff}$ around an optimal value associated with each opacity peak;  the optimal $T_{\rm eff}$ is higher if the temperature at the opacity peak is higher. 
The Fe opacity peak at $T\approx 2\times10^5$\,K is responsible for pulsations of B-type stars, while the peak at He II ionization zone ($T\sim 4\times 10^4$\,K)  is responsible for $\delta$ Sct and the classical Cepheids.   

If the luminosity to mass ratio is very large; i.e., $L/M \gtrsim 10^4 L_\odot/M_\odot$, the $\kappa$-mechanism is enhanced by the confinement of pulsation energy around an opacity peak, and the strange-mode instability occurs. 
This corresponds to the nearly horizontal boundary in Fig.\,\ref{fig:allvar}, which bounds the distribution of $\alpha$ Cygni variables. 

In the following, we discuss excitations of A- and B-type variables in direction of decreasing luminosity.

\section{$\alpha$ Cygni variables; AB-type bright supergiants}

The $\alpha$ Cygni variables show semi-regular light and radial-velocity variations with periods of roughly 10 to $10^2$ days.
At least some of the semi-periodic variations are considered to be caused by stellar pulsations.
Some of the $\alpha$ Cygni variables are LBVs (S Dor variables) showing long-term (a few years to decades) variations, on which $\alpha$-Cygni type micro variations are superposed.
As seen in Fig.,\ref{fig:allvar}, the distribution boundary of $\alpha$ Cygni variables is roughly horizontal ($\log L/L_\odot \gtrsim 4.7$), involving massive stars whose initial masses are larger than about $20\,M_\odot$.
The horizontal boundary indicates that the strange mode instability (Glatzel\cite{gla94},Saio et al.\cite{sai98}, Saio\cite{sai09}) is responsible to excite the pulsations of these stars.

For normal blue to red evolutions of massive stars, however, the instability boundary occurs at $\log L/L_\odot\approx 5.9$ (\cite{sai09}) corresponding a initial mass larger than $50\,M_\odot$; the boundary is far too luminous compared with  the observational  boundary of the distribution of $\alpha$ Cygni variables in the HR diagram.
The problem can be solved by assuming that the $\alpha$ Cygni variables are evolving bluewards after losing significant mass in the red-supergiant phase (Saio et al.\cite{sgm13}). 
However, these models predict surface N/C and N/O ratios which are much larger than the results of spectroscopic analyses (e.g.,  Przybilla et al.\cite{pri10}).
Since the surface chemical compositions depend on the internal mixing, the solution of the problem  
might constrain the assumptions on internal mixing during the evolution of massive stars.

\section{B-type variable stars}

\subsection{Excitation}
Pulsations of B-type stars are excited around the Fe opacity peak at $T\sim 2\times10^5$K. %(\cite{rog92}).
P-mode ($\beta$ Cephei) instability region (dash-dotted line in Fig.\,\ref{fig:allvar}) is located on the hotter side of the two g-mode (SPB) instability regions (solid lines), although these excitations are caused by the same opacity peak.
This difference is interpreted by the relation between the thermal time in the driving zone (i.e., opacity peak) and the pulsation period.
As mentioned in \S\ref{sec:intro}, $\kappa$-mechanism driving is most effective if the thermal timescale is comparable to the pulsation period, and the thermal timescale at a certain temperature zone is shorter in the hotter model.
For this reason, a shorter-period p-mode instability region is hotter than that of g-modes with longer periods.

\begin{figure}[!t]
\begin{center}
 \includegraphics[width=0.55\textwidth]{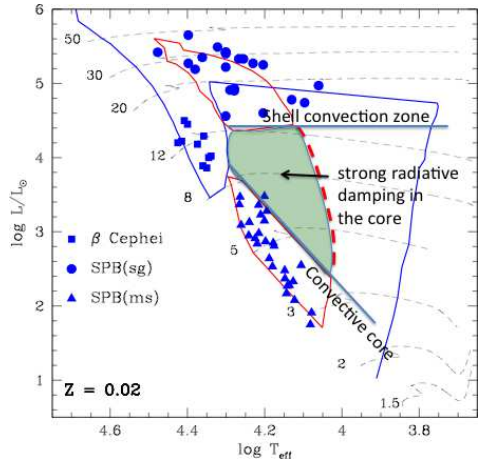}
\vspace{-3mm}
\caption[]{The instability regions for p-modes ($\beta$ Cephei variables) and for g-modes (slowly pulsating B (SPB) stars).
The g-mode instability region is separated into two regions by the effect of strong radiative damping in the deep interior.
One of the boundary of the interruption corresponds to the end of main-sequence phase (i.e., disappearance of convective core), while the other one corresponds to the lower luminosity bound for the appearance of the intermediate convection zone above the hydrogen burning shell.
}
\label{fig:gmodes}
\end{center}
\end{figure}

The g-mode instability region itself is separated into two regions. 
If the instability were determined only by the  efficiency of the $\kappa$ mechanism, the separation would not occur as indicated by a dashed line in Fig.\,\ref{fig:gmodes}.
The interruption arises from a change in the internal structure of the post-main-sequence stars, by which the g-mode property is affected strongly.
As a star terminates the main-sequence evolution, a convective core is replaced with a radiative dense core having a very high Brunt-V\"ais\"al\"a frequency $N$, which is defined as
\begin{equation}
N^2 = g\left({1\over\Gamma_1}{d\ln p\over dr}- {d\ln\rho\over dr}\right),
\end{equation}
where $g$ is the local gravitational acceleration, $p$ pressure, $\rho$ density, and $r$ distance from the center.
The typical wavelength of a g-mode being $\propto r\sigma/N$ (e.g., \cite{aer10,unn89}; $\sigma=$ g-mode frequency) is very short in the radiative core having very high $N$; the short wavelength enhances radiative damping there.
For this reason, g-modes are generally expected to be damped in post main-sequence stars, which generates the instability boundary at the termination of main-sequence phase for stars with initial masses of $\lesssim 11M_\odot$ (Fig.\,\ref{fig:gmodes}).
During the post main-sequence phase of a more massive star, however, a shell convection zone appears above a hydrogen burning shell; the shell convection zone can reflect g-modes and prevent g-modes from suffering strong radiative damping in the interior (e.g., Saio et al.\cite{sai06}).
The reflection makes g-mode excitation possible above the luminosity boundary at $\log L/L_\odot \approx  4.4$ ($M\approx 12\,M_\odot$) in the post-main-sequence stage.  
This explains periodic light variations in blue supergiants found by Hipparcos and MOST satellites (\cite{sai06}, Lefever et al.\cite{lef07}). 
If the luminosity boundary is accurately determined  observationally, it will provide a test for evolutionary models. 
The effects of overshooting and mass loss on the appearance of the shell convection zone is discussed in Godart et al.\cite{god09}.

The distribution boundary of the main-sequence SPB stars tells us the exact locus of the termination of main-sequence phase which depends on the assumption of core overshooting. 
The distribution shown in Fig.\,\ref{fig:gmodes} seems to be consistent with the theoretical prediction by models without core overshooting, indicating that little core-overshooting occurs in the main-sequence SPB stars (they are known to be relatively slow rotators).
This contradicts the recent estimate of a 0.2--0.3$H_p$ ($H_p=$ pressure scale height) core-overshoot from period spacing of the SPB star HD 50230 by Degroote et al.\cite{deg10}. 
Furthermore, detailed comparisons with observed frequencies of $\beta$ Cephei stars tend to indicate the presence of substantial ($0.1-0.3\,H_p$) core-overshooting (see discussion in e.g. \cite{sai12}). 

Figure\,\ref{fig:allvar} also shows the positions of rapidly rotating Be stars, most of which are known to show short-term light and spectroscopic variations attributed to g-mode pulsations (sometimes called SPBe). 
The distribution extends well beyond the theoretical main-sequence band for models without core-overshooting.
This indicates the presence of  considerable internal mixing in these rapidly rotating stars, which prolongs the main-sequence evolution.
The required mixing corresponds to a core-overshooting up to $\sim 0.35H_p$ (Neiner et al.\cite{nei12}).

\subsection{Observed frequency distributions of g-modes in rapidly rotating stars}
In a rapidly rotating star, frequencies in the co-rotating frame, $\sigma_{\rm g}$,  of intermediate to high order g-modes are much smaller than $|m\Omega|$ ( i.e., $\sigma_{\rm g}\ll |m\Omega|$),  where $m$ and $\Omega$ are azimuthal order of the nonradial pulsation and the angular frequency of rotation (uniform rotation assumed), respectively.
Then, the frequencies in the observer's frame are $\sigma_{\rm g}-m\Omega \approx -m\Omega$; they form groups associated with azimuthal-order $m$ (here we adopt the convention that $m<0$ corresponds to prograde modes).  
Such frequency groupings have been detected in the Be star HD 163868 by a MOST observation (Walker et al.\cite{wal05}), in which two frequency groups  are identified as prograde sectoral g-modes of order $m=-1$ and $-2$. 
This identification is consistent with the theoretical prediction that prograde sectoral modes are least affected by damping due to mode-couplings and likely excited (Aprilia et al.\cite{apr11}).
The property can be used to determine the rotation frequencies of Be stars without relying on the spectroscopic information of rotation velocities (see Cameron et al.\cite{cam08}, Saio\cite{sai13}). 
We note that the frequency groupings of g-modes can also occur in rapidly rotating normal B stars as found by Kepler (Balona et al.\cite{bal11}).

\section{Main-sequence A-type variable stars}

\begin{figure}[!t]
\begin{center}
 \includegraphics[width=0.55\textwidth]{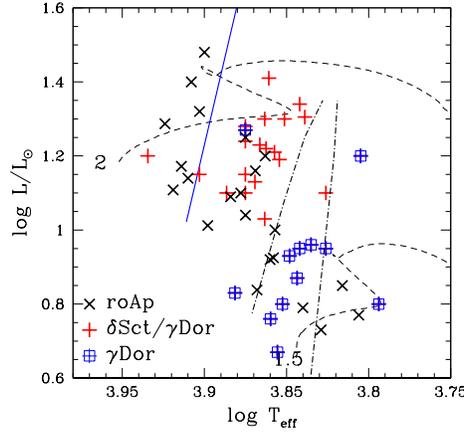}
\vspace{-5mm}
\caption[]{Distribution of A-type main-sequence variables, $\delta$ Sct, $\gamma$ Dor variables, and roAp stars on the HR diagram. 
Solid line indicates an approximate position of the blue edge of low-order radial pulsations.
The range bounded by dot-dashed lines is the instability region of g-modes (i.e., $\gamma$ Dor stars) obtained by Dupret et al. \cite{dup05} for a mixing-length of $2\,H_p$ taking into account the interaction between pulsation and convection.
}
\label{fig:Avari}
\end{center}
\end{figure}

\begin{figure}[!t]
\begin{center}
 \includegraphics[width=0.55\textwidth]{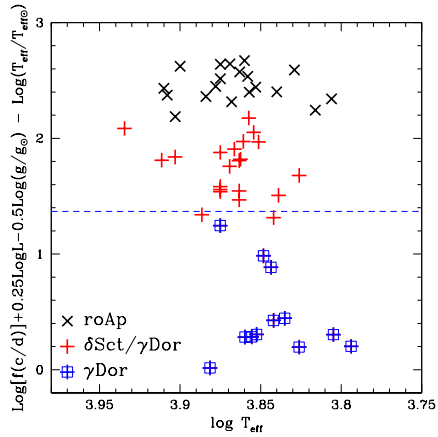}
\vspace{-7mm}
\caption[]{
Frequencies normalized with respect to the sun for the stars shown in Fig.\,\ref{fig:Avari}.
 A horizontal dashed line indicates approximately the frequency of the fundamental radial pulsation. Although the three types of variables are distributed similarly on the HR diagram (Fig.\,\ref{fig:Avari}), they have distinctly different frequencies, corresponding to g-modes ($\gamma$ Dor variables), low-order p-modes ($\delta$ Sct variables), and high-order p-modes (roAp stars).}
\label{fig:Avari_freq}
\end{center}
\end{figure}

The main-sequence A-type variables; $\delta$ Sct, $\gamma$ Dor, and roAp stars are located approximately in the same region of the HRD, where the Cepheid instability strip crosses the main sequence (Figs.\,\ref{fig:allvar}, \ref{fig:Avari}).
These groups pulsate in different ranges of frequencies (or periods) as seen Fig\,\ref{fig:Avari_freq}, which shows the dominant frequency of each star normalized to the sun (to eliminate the effects of different luminosity and effective temperature). 
The horizontal dashed line indicates the frequency of radial fundamental mode. 
As we see in this figure, g-modes, low-order p-modes, and high-order p-modes  are excited in $\gamma$ Dor stars, $\delta$ Sct stas, and in roAp stars, respectively. 

As is well known, the p-mode pulsations of $\delta$ Sct stars are excited by the $\kappa$-mechanism in the He II ionization zone.
In a broad sense, the relation between the optimal $T_{\rm eff}$ for the p-mode excitation in $\delta$ Sct stars and the g-mode excitation in $\gamma$ Dor stars by the He II opacity peak is similar to the relation between p-modes of $\beta$ Cep stars and g-modes in SPB stars by the Fe opacity peak.
The former case, however, is more complicated because in the He II ionization zone  of $\gamma$ Dor stars efficient convection occurs, while Fe opacity peak region (for the latter case) is radiative in main-sequence SPB stars.
Therefore, it is necessary to include convection-pulsation couplings in order to  examin the stability of g-modes in $\gamma$ Dor stars.
The region bounded by the two dash-dotted lines in Fig.\,\ref{fig:Avari} is a region predicted for g-modes to be excited  (Dupret et al.\cite{dup05}).
The  range is roughly consistent with the distribution of $\gamma$ Dor stars.

However, recent space photometries by Kepler and CoRoT satellites revealed that many $\delta$ Sct stars show g-mode pulsations as well; i.e., there are many $\delta$ Sct - $\gamma$ Dor hybrids (see Grigahc\`ene et al.\cite{gri10} for detail). 
Therefore, at present we do not understand the excitation of g-modes in A-type main-sequence stars.

\subsection{Excitation of high-order p-modes in roAp stars}

\begin{figure}[!t]
\begin{center}
 \includegraphics[width=0.55\textwidth]{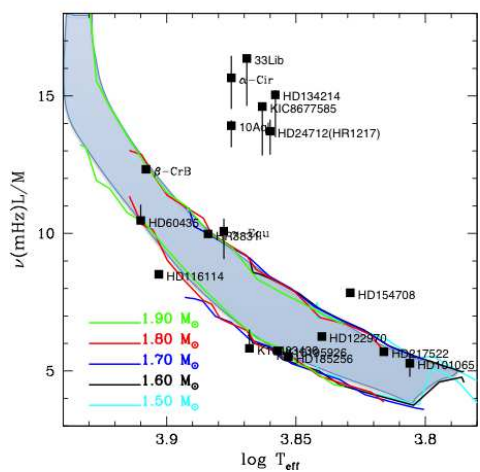}
\vspace{-3mm}
\caption[]{
The range of high-order p-modes being excited (shaded region) and some of the observed roAp stars are plotted for comparison. 
}
\label{fig:roap_stbil}
\end{center}
\end{figure}

As we see in Fig.\,\ref{fig:Avari_freq}, roAp stars pulsate in high-order p-modes, comparable to the solar 5\,min oscillations.
Although roAp stars have effective temperatures similar to those of $\delta$ Sct stars, He II ionization zone is ineffective to drive such high order p-modes because the periods are much shorter than the thermal timescale there.
Rather, the periods are comparable to the timescale in H-ionization zone, but the driving effect is weakened by the convective flux.
Balmforth et al. \cite{bal01} found that high order p-modes are excited by the $\kappa$-mechanism in the H-ionization zone if a strong magnetic field suppresses convection in the polar region.   
The excitation region on the HR diagram (Cunha \cite{cun02}) seems consistent with most of the roAp strs.
However, some inconsistency becomes apparent if the frequency ranges of  roAp stars are compared.

Figure\,\ref{fig:roap_stbil} shows the instability region of high-order p-modes (shaded region) in the $\log T_{\rm eff} - \nu L/M$ plane, where $\nu$ is oscillation frequency, $L$ and $M$ are in solar units.
In this diagram, minimum and maximum frequencies of high-order p-modes excited in various models (with suppressed convection) are located along two lines, respectively, forming the instability region. 
Also plotted are selected roAp stars having relatively well determined parameters.
Obviously, some well studied roAp stars are located far from the instability region.
They are above the acoustic cut-off frequency. 
Currently, it is not clear how to excite these modes, the effect of atmospheric temperature inversions considered in Gautschy et al. \cite{gau98} seems not strong enough.

\bigskip
{\it Acknowledgements.} I am grateful to Alfred Gautschy for useful comments.